\documentclass[
preprint
,tightenlines
,showpacs
,showkeys
,nofootinbib
,aps
,amsfonts
,amssymb
,byrevtex
,pdftex  
]{revtex4-1}

\usepackage{graphicx}  
\usepackage{amsmath, amssymb, bm}   
\usepackage[dvips]{color}
\usepackage{pslatex}  
\usepackage{natbib}
\usepackage{hyperref}

\def\r{{\it r}}

\def\){\right)} 
\def\({\left(} 
\def\]{\right]} 
\def\[{\left[}

\begin{document}

\title{
Constraining phases of quark matter with studies of \r-mode damping in neutron stars
}

\author{%
Gautam Rupak}
\email{grupak@u.washington.edu}
\affiliation{Department of Physics $\&$ Astronomy,\\
 Mississippi State University, Mississippi State, MS 39762, U.S.A.}
\author{%
Prashanth Jaikumar
}
\email{pjaikuma@csulb.edu}
\affiliation{California State University Long Beach, Long Beach, CA
  90840, U.S.A. }
 \affiliation{Institute of Mathematical Sciences, CIT Campus, Chennai 600113, India}

\begin{abstract}
The \r-mode instability in rotating compact stars is used to 
constrain the phase of matter at high density. 
The color-flavor-locked phase
with kaon condensation (CFL-K0) and without (CFL) is considered in the 
temperature range $10^8$K $\lesssim T\lesssim 10^{11}$K. 
While the bulk viscosity
in either phase is only effective at damping the \r-mode at temperatures 
$T\gtrsim10^{11}$K, the shear viscosity in the CFL-K0 phase is the only effective
damping agent  all the way down to temperatures 
$T\gtrsim 10^8$K characteristic of cooling neutron stars. 
However, it  cannot keep the star from becoming unstable to gravitational wave emission for rotation frequencies 
$\nu\approx 56-11$Hz at $T\approx10^8-10^9$K.  
Stars composed almost entirely of CFL or CFL-K0 matter are 
ruled out by observation of rapidly rotating neutron stars, indicating that
dissipation at the quark-hadron interface or nuclear crust interface must 
play a key role in damping the instability.

\end{abstract}



\maketitle

\section{\label{sec:Intro}Introduction}

 The $r$-mode instability of rotating compact stars
is of wide astrophysical interest for several
reasons~\cite{Andersson:1997xt,*Friedman:1997uh,*Lindblom:1998wf}. The majority of neutron stars
have rotation period $P\sim 0.5$seconds, much slower than theoretical
limits: $P\sim 0.001$s ($\nu=1/P\sim 1$kHz)~\cite{Shapiro:1984}. Even the fastest
spinning neutron star PSR J1748-2446ad with $\nu=716$Hz~\cite{Hessels:2006} is not near the theoretical stability
limit, with the majority of  ``spun-up'' neutron stars in low mass X-ray binaries (LMXBs)
catalogued between 300-640 Hz. 
The \r-mode provides a mechanism for spinning down
young neutron stars~\cite{Andersson:1998qs} and limiting spin-up
frequencies in millisecond pulsars (see eg.~\cite{Lorimer:2008se} for a
review) through angular momentum loss in gravitational wave emissions. 
These gravitational waves
may be detected given the expected improved sensitivities of 
ground-based interferometers eg. VIRGO, advanced
LIGO~\cite{Sa:2007zz,*Watts:2008qw,*Stergioulas:lrr-2003-3}. 
In addition, \r-modes are probes of the
equation of state of cold and dense matter  through viscous damping effects and advance
our understanding of nuclear and quark interactions by providing an
astrophysical and phenomenological link to Quantum Chromodynamics
(QCD) over a certain range of temperatures ($10^7$K $\lesssim T\lesssim10^{11}$K) and
at super-saturation densities [$n_B\sim$ (2-5)$n_{\rm sat}$]. Only a few
studies of \r-modes linking them to phenomenology of quark matter in
neutron stars have been carried out thus
far~\cite{Madsen:1999ci,Andersson:2001ev,Jaikumar:2008kh,Sa'd:2008gf,Mannarelli:2008je}. 
In this work we  
use \r-mode damping to constrain the presence of a
realistic phase of superconducting quark matter~\cite{Alford:1998mk,Rapp:1997zu} in the core of neutron stars: viz.,
one that includes the effects of a sizable
strange quark mass $m_s$ at intermediate baryon density. 

A key parameter in understanding the \r-mode instability in rotating
compact stars is the critical frequency $\Omega_c$ for the onset of
the instability ($\Omega=2\pi/P$~rad/s).  Gravitational radiation tends to make the $r$-mode
grow on a characteristic time scale $\tau_{GR}$. Internal friction
such as viscosity or mutual friction tends to damp $r$-mode growth on
a characteristic time scale $\tau_F$. The competition between the two
determines $\Omega_c$ at any temperature. At $\Omega\geq\Omega_c$, the
$r$-mode develops an instability ($1/\tau_{\rm total}=1/\tau_{GR}+1/\tau_{F}<0$). 
Previous works~\cite{Sa'd:2008gf,Jaikumar:2008kh,Mannarelli:2008je} have studied the
\r-mode damping in normal (ungapped) quark matter as well as some
superconducting (gapped) phases such as the color-flavor-locked phase
(CFL), and found that bulk and shear viscosity in the  fully
gapped phase is large enough to render rapidly spinning quark stars
($\Omega\geq 0.1\Omega_\mathrm{Kepler}$) stable.  However, this conclusion applies
only in a restricted temperature regime $T\gtrsim 10^{10}$K, since
the shearing mean free path due to phonons otherwise becomes larger
than the star's radius ($\sim 10$km). Furthermore, for temperatures
$T\lesssim 10^8$K, mutual friction in the CFL phase has been shown to be
too weak to damp \r-modes~\cite{Mannarelli:2008je}, arguing for rapid spin-down of
CFL stars to less than a few Hz, effectively ruling our cold CFL
matter in any rapidly rotating neutron stars. The \r-mode damping in
the temperature regime $10^8$K $\lesssim T \lesssim10^{10}$K for color-superconducting
phases of quark matter is therefore an open question, and is the
focus of this work. Here we study $r$-mode damping and
the critical frequency curve for a neutron star made up mostly or
entirely of the kaon condensed CFL (CFL-K0) phase. 

The CFL phase with symmetric pairing of up-, down- and strange-quarks is severely stressed
at realistic chemical potential $\mu_q\sim 300$ MeV 
because the strange quark mass $m_s\sim 100$ MeV is
non-negligible and costs extra energy $m_s^2/(2\mu_q)$ compared  to the much lighter up- and
down-quarks. In Ref.~\cite{Bedaque:2001je}, it was shown that it is energetically favorable for
the CFL vacuum to be in a rotated chiral state with a non-zero kaon
condensate when the cost $m_s^2/(2\mu_q)$ exceeds the kaon mass
$m_K$, the lightest meson in high-density
QCD~\cite{Son:1999cm}. Realistic neutron star densities could support
the CFL-K0 phase. 
We find that in the CFL-K0 phase, while the bulk viscosity has little role to play in the temperature region relevant to \r-mode damping in cooling neutron
stars, the shear viscosity from kaons in the CFL-K0 phase is important
even below $T\lesssim 10^9$K, unlike the CFL phase. In essence, this fact
coupled with the large mean free path of the phonons controls the main
features of the critical frequency curve (Fig.~\ref{fig:critfreq}).

\section{\label{sec:rmode}The $r$-mode}

\r-mode is a pulsation mode 
intimately linked to the rotational properties of the star, and 
the restoring force here is the Coriolis force. The frequency of $r$-mode oscillations depends on the 
star's rotation frequency as well as the average density 
of the star. Including second-order rotational effects,
 the mode 
frequency $\omega_r$ in the inertial frame is ~\cite{Provost:1981,Lindblom:1998wf}
\begin{align}
\label{kappa2} 
\omega_r &=  \omega_{rot}-m\Omega
\approx \[\frac{2}{m+1}-m\]\Omega +\kappa_2\frac{\Omega^2}{\pi G\bar{\rho}_0}
+\mathcal O(\Omega^4), 
\end{align}
where $\omega_{rot}$ is the mode frequency in the co-rotating frame and
$m$ is the azimuthal quantum number of the mode, taken to be 2 for
the first \r-mode to become unstable (higher $m$ modes become unstable
at higher frequency). 
The number $\kappa_2$ includes the sensitivity to the density profile
(see Refs.~\cite{Lindblom:1999yk,Jaikumar:2008kh}), 
$\bar{\rho}_0$ is the average density of the unperturbed star and $G$
is Newton's constant.

The \r-mode couples to gravitational waves through the current multipole
of the perturbation~\cite{Lindblom:1998wf}. All modes with azimuthal quantum number $m\geq 2$ suffer the so-called CFS instability~\cite{Chandrasekhar:1992pr,Friedman:1978hf} when $\Omega>\omega_{rot}/m$ so that the $r$-mode energy
grows at the expense of the star's rotational kinetic energy as angular momentum is lost to gravitational wave emission. In general, viscous forces counter this energy growth. The timescale $\tau$ associated with growth or dissipation is given by
$1/{\tau_i}=-\frac{1}{2 E}\left(\frac{dE}{dt}\right)_i$, 
where the subscript $i$ corresponds to various dissipative forces. 
Details about computing $E$ and $\left(\frac{d E}{d t}\right)_{i}$
  are given in Refs.~\cite{Lindblom:1999yk,Jaikumar:2008kh}.
Explicitly, the gravitational radiation timescale 
is~\cite{Lindblom:1998wf}
\begin{align}
\label{taugw}
\frac{1}{\tau_{GW}} = - \frac{32 \pi G}{c} 
\frac{\left(m-1\right)^{2m}}{\left[\left(2m+1\right)!!\right]^2} 
\int_0^R dr\rho\[r \frac{\Omega}{c}
  \frac{m+2}{m+1}\]^{2m+2} .
\end{align}
This timescale is negative, indicating exponential
mode growth (the instability).

Viscosity (bulk, shear and otherwise) is a source of damping for the \r-mode. 
For a given phase with shear viscosity $\eta$, the shear viscosity
timescale 
is~\cite{Lindblom:1998wf}
\begin{align}\label{InvShear}
\frac{1}{\tau_{\eta}} = \frac{(m-1)(2 m+1)}{\int_0^R dr \rho r^{2m+2}}
\int_0^R dr\ \eta\ r^{2m} .
\end{align}
We find that the bulk viscosity
  damping timescale plays a minor role in our calculation, see
  Table~\ref{timetable} and
  Figs.~\ref{fig:Viscosities},~\ref{fig:critfreq}.

Mutual friction is another source of dissipation for the \r-mode energy.
It occurs due to the scattering of phonons from quantized vortices in the 
superfluid component. For the CFL phase, this has been calculated to be~\cite{Mannarelli:2008je}:
$\frac{1}{\tau_{MF}}\simeq 36.2 \left(\frac{T}{\mu_q}\right)^5\Omega$
in the limit of negligible vortex mass. 
The critical rotation frequency $\Omega_c$ of
the star can be determined by the criterion that
at this frequency, the fraction of energy dissipated per unit time
exactly cancels the fraction of energy fed into the $r$-mode by
gravitational wave emission:
\begin{align}
\label{criteqn} \frac{1}{\tau_{\rm total}}\Big|_{\Omega_c}\equiv\left[\frac{1}{\tau_{\rm GW}}+ \frac{1}{\tau_{\eta}}+\frac{1}{\tau_{\zeta}}+\frac{1}{\tau_{\rm MF}}\right]\Big|_{\Omega_c}=0 . 
\end{align}

Stable rotation frequencies at any temperature are limited by the
smaller of the critical frequency as computed from Eq.~(\ref{criteqn}) or the Kepler limit
$\Omega/\Omega_K$=1.  As depicted in Fig.~\ref{fig:critfreq}, the
region above the temperature-dependent $\Omega_c$ curve is unstable to
\r-mode oscillations and the star upon entering this region will be
spun down rapidly to $\Omega<\Omega_c$ by emitting gravitational waves
and losing its angular momentum in the process.

\section{\label{sec:viscosity}CFL-K0: EoS and Viscosity}

For
simplicity, we adopt a homogeneous composition assuming that the
entire star is  made of CFL-K0 matter.  This allows a direct comparison with \r-mode
calculations in hadronic matter with and without a crust~\cite{Bildsten:1999zn,*Levin:2000vq}.
We will see later that a crust is   
required for agreement with observations, but the more elementary case
of a pure CFL-K0 star  
should be tested first.
Just as for the CFL phase, the neutral kaons are expected
to dominate over the phonons in determining the bulk
viscosity. However, the condensation changes the temperature
dependence of the viscosity, which is the most important outcome for
the \r-mode damping. The pressure of the CFL-K0 phase, including terms up to ${\cal
  O}(m_s^4)$ is given by~\cite{Bedaque:2001je,Forbes:2004ww}  
\begin{align}
\label{Pcflc0}
P_{\rm CFL-K0} = P_{\rm CFL} +\frac{1}{2}f_\pi^2\(\frac{m_s^2}{2k_f}\)^2,
\end{align}
The pressure from the Nambu-Goldstone bosons introduces a negligible temperature
dependence. 
We use the perturbative expression for $f_{\pi}$~\cite{Son:1999cm}. 
This EoS differs
from the CFL EoS only at ${\cal O}(m_s^4)$. In the above expression,
the common Fermi momenta $k_F$ also determines the CFL pressure~\cite{Alford:2001zr}
\begin{align}
\label{Pcfl}
P_{\rm CFL} =-\frac{3}{\pi^2}\sum_{i=u,d,s}\int_0^{k_F}dk\
k^2\[\sqrt{k^2+m_i^2}-\mu\]
+\frac{3\Delta^2\mu^2}{\pi^2} - B ,
\end{align}
where $\Delta$ is the superconducting quark gap and $B$ is the MIT Bag 
constant. We choose $\Delta=100$ MeV  and $B=80$ MeV/fm$^3$ such that the energy per baryon
$E/A < 930$ MeV at zero pressure. 
In the 
limit of vanishing up and down quark mass, we have up to 
${\cal O}(m_s^4)$: 
$k_F=\frac{1}{3} \left(6 \mu -\sqrt{9 \mu
    ^2+ 3m_s^2}\right)\approx
\mu-{m_s^2}/({6\mu})+{m_s^4}/({72\mu^3})$.

Evaluating the expression for $P_{\rm CFL-K0}$ pressure using
Eq.(\ref{Pcfl}), we obtain
\begin{align}
P_{\rm CFL-K0}\approx&\frac{3\mu^4}{4\pi^2}-B+\frac{3\mu^2}{4\pi^2}\(4\Delta^2-m_s^2\)
+\frac{m_s^4}{32\pi^2}\(1+6\ln\frac{4\mu^2}{m_s^2} +
4\pi^2\bar{f}_\pi^2\), 
\end{align}
where
$\bar{f}_\pi^2 \equiv{f_\pi^2}/{k_F^2}
\approx
({21-8\ln2})/({36\pi^2})$ .
The energy density then follows from standard 
thermodynamic relation
$\epsilon= -P_{\rm CFL-K0}+ n\mu$ with 
 number density $n=\partial_\mu P_{\rm CFL-K0}$.
Eliminating $\mu$ between pressure and energy density, we get the EoS 
in the useful form
\begin{align}
P_{\rm CFL-K0}\approx
\frac{1}{3}\(\epsilon-4B\)+\frac{4\Delta^2-m_s^2}{3\pi}\sqrt{\epsilon-B}
-\frac{m_s^4}{12\pi^2}\[1+
\(1-\frac{4\Delta^2}{m_s^2}\)^2 -2\pi^2\bar{f}_\pi^2
\right. \\
\left. 
-3\ln\frac{8\pi\sqrt{\epsilon-B}}{3m_s^2}\] .\nonumber
\end{align}

\begin{figure}[tbh]
\begin{center}
\includegraphics[width=0.65\textwidth,clip=true]{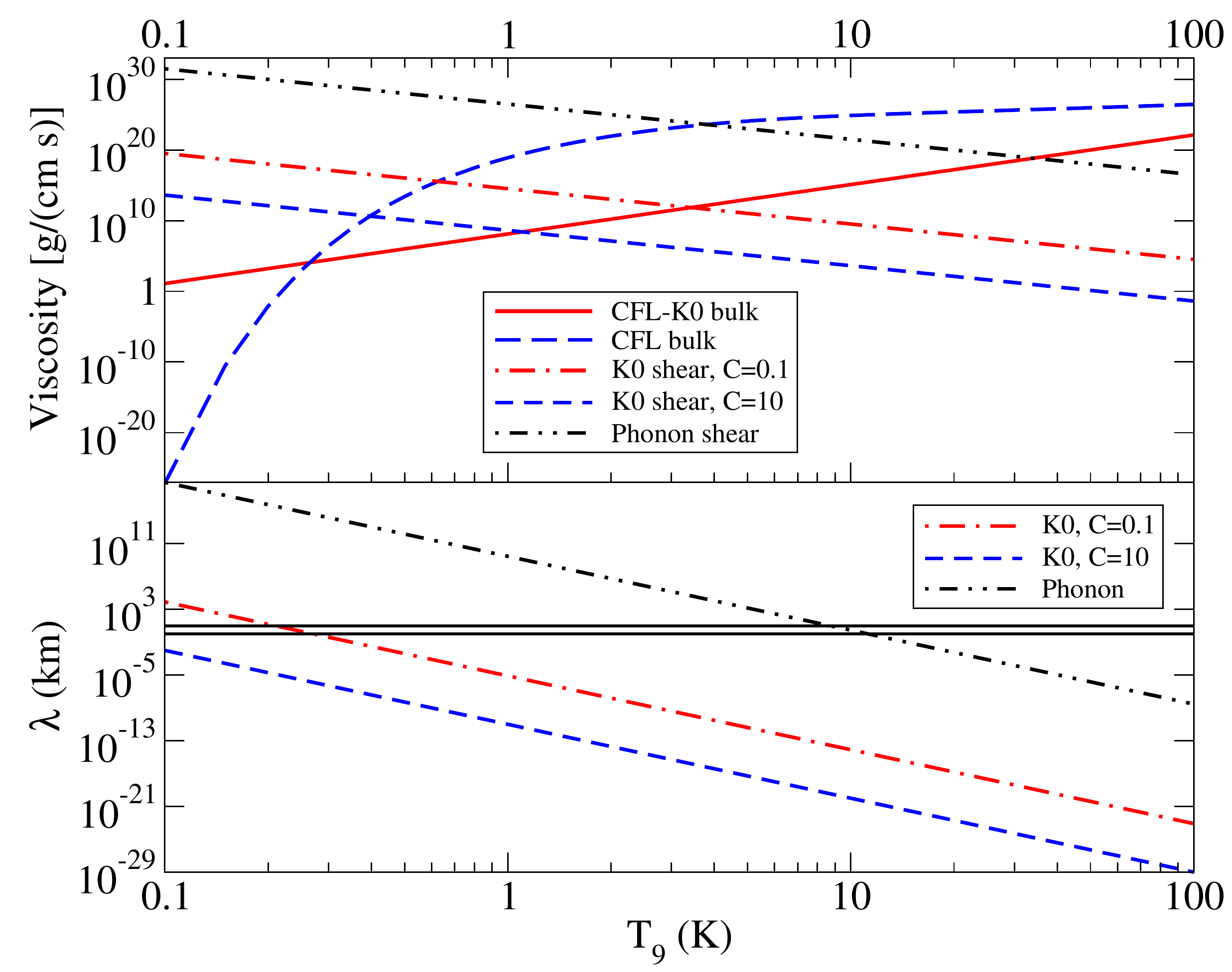}
\end{center}
\caption{ \protect 
Top panel: viscosities for the CFL and CFL-K0 phase. 
Bottom panel: phonon and kaon shear viscosity mean free path
  $\lambda$. 
 The typical parameter values~\cite{Alford:2007rw,Jaikumar:2008kh} are $\mu_q=310$
  MeV, $\mu_k=17.92$ MeV and $\omega= 2\pi$
ms$^{-1}$, $\delta m=1$
MeV for CFL,  $\delta m=-1$
MeV for CFL-K0. In the bottom panel, the crossings with the solid (black) horizontal lines  at 1km and
  10km indicate
when $\lambda$ is on the order of the stellar size, making
the shear viscosity irrelevant.
}
\label{fig:Viscosities}
\end{figure}

\begin{figure}[tbh]
\begin{center}
\includegraphics[width=0.65\textwidth,clip=true]{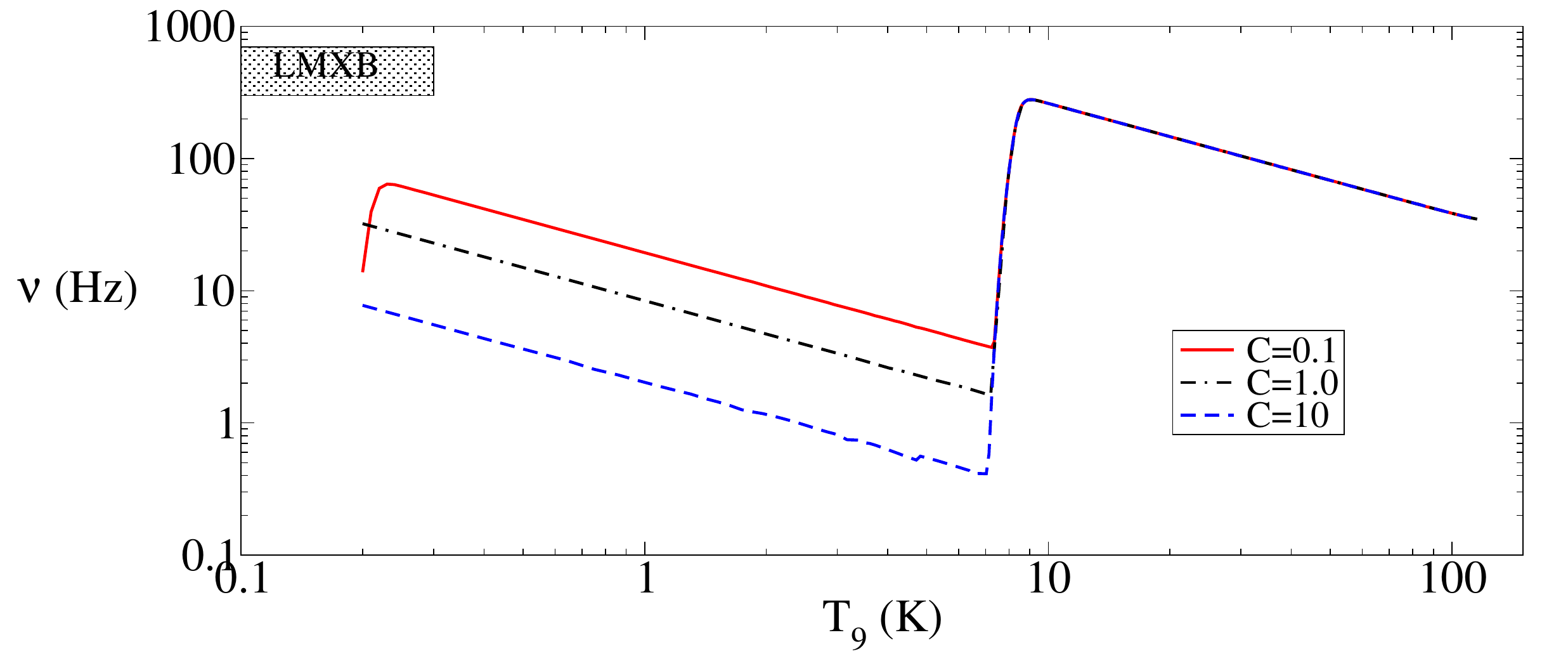}
\caption{The critical frequency $\nu$  
 as a function of
 temperature for CFL-K0 quark stars. The curves end abruptly when the
 shear mean free path exceeds the stellar radius. We use the same
 parameters as in Fig.~\ref{fig:Viscosities}. 
 The box represents typical temperatures~\cite{Brown:2002rf} 
  ($2 \times 10^{7}$-$3 \times 10^8$ K) and rotation
frequencies (300-700~Hz) of the majority of observed neutron stars in LMXBs.}
\label{fig:critfreq}
\end{center}
\end{figure}

The dominant contribution to the bulk viscosity in the fully gapped 
CFL or CFL-K0 phase  is determined by weak reactions involving the
lightest modes~\cite{Alford:2007rw,Alford:2008pb}. 
In the CFL-K0 phase, this is the neutral kaons, a fraction
of which are in the condensate. 
From Alford et. al~\cite{Alford:2008pb}, bulk viscosity is 
 \begin{align}
\label{bulkv}
\zeta=\frac{n^2}{\chi^2}\frac{\lambda}{\omega^2+\lambda^2/\chi^2},
\end{align}
where the kaon number density $n$, susceptibility $\chi$ and weak
equilibration rate for kaons in the condensate are computed in
Ref.~\cite{Alford:2008pb} using 
the 2PI formalism in order to assess thermal effects consistently.
Note that $\delta m = m_K-\mu_K<0$ for the condensed
 phase but $\delta m >0$ for the pure CFL phase. For the 
typical parameter values~\cite{Alford:2007rw,Jaikumar:2008kh} used 
in Fig.~\ref{fig:Viscosities}, $T_c\sim 3\times 10^{11}-4\times
 10^{11}$ K. There is a negligible contribution to bulk viscosity due
 to phonon scattering above $T\gtrsim 5\times 10^8$K~\cite{Manuel:2007pz}.

Kaon condensation  breaks flavor symmetry, and the associated Nambu-Goldstone mode contributes to shear viscosity at low temperature.  Although the kaonic contribution is smaller  than that from the
phonons, it becomes important when the phonon mean free path exceeds the
stellar radius at $T\lesssim 10^{10}$K - see Fig~\ref{fig:Viscosities}. 
The leading order contribution  was calculated in terms of an unknown dimensionless coefficient $C$~\cite{Alford:2009jm}.  Analytic expressions for the shear viscosity could be derived for very small or large value of the coupling $C$:
\begin{align}
\label{shearv}
C\ll 1:\ \  \eta& 
\approx
3.44\times 10^{-4} \frac{\nu^{11}}{C^2}\frac{f_\pi^4\Delta^4}{T^5}, \\
C\gg 1:\ \ \eta& 
\approx 
1.7\times 10^{-8} \frac{\nu^{11}}{C^4}
\frac{f_\pi^4\Delta^8}{\sin^4(\phi/f_\pi)\mu_K^4 T^5} \nonumber.
\end{align}
For natural values $C\sim 1$, shear viscosity $\eta$ has to be calculated numerically which we have done and cross-checked
against the results of Ref.~\cite{Alford:2009jm}.  Here also $\eta$ shows a $T^{-5}$ dependence similar to the contribution from massless phonon scattering. Phonon contribution to shear viscosity is
\begin{align}
\eta_\mathrm{Phonon} = 3.745\times 10^{6} \(\frac{\mu_q}{\rm MeV}\)^8 T_9^{-5} \mathrm{ g/(cm\ s)} \,,
\end{align}
where $T_9$ is the temperature in units of $10^9$K. The effects of mutual friction have been shown to be too small to
effectively damp the \r-mode instability at $T\lesssim 10^8$K~\cite{Mannarelli:2008je}. In this work,
we include the mutual friction damping.

In Fig.~\ref{fig:Viscosities} below, we show the bulk and shear viscosity from the CFL-K0 phase in comparison to the pure CFL phase.

\section{\label{damping}Damping effects}

Next, we present results for the damping times from
the various mechanisms discussed previously and obtain the
critical frequency plot. The damping times are listed in Table~\ref{timetable}.
\begin{table}[tbh]
\begin{center}
\caption{Damping times at $T=10^9$K for the CFL-K0 phase, 
the CFL phase {\it with} ${\cal O}(m_s^4)$ corrections (labeled CFL4 ) and {\it without}. 
For the CFL-K0 phase, the shear damping timescale due to kaons with
parameter value $C=1$  (see text) is shown. 
The central energy density ${\rho}_c$ (in units of 
nuclear saturation density $\rho_0=2.54\times 10^{14}
\mathrm{g/cm^3}$) 
is chosen to yield a 1.4 
$\mathrm{M}_{\odot}$ star. The stellar rotation frequency 
is taken to be half the Kepler frequency
$\Omega_K=\frac{4}{9}\sqrt{2\pi G\bar{\rho}}$.
}
\label{timetable}
\begin{ruledtabular}
\begin{tabular}{lcccccc}
EoS  & 
${\rho}_c/\rho_0$ & $R$ (km) &
$\Omega_K$ (rad/s) & $\tau_{\mathrm{\zeta}}$ (s) &
$\tau_{\mathrm{\eta}}$ (s)
& $\tau_{\mathrm{GW}}$ (s) \\
\hline
CFL-K0 & 2.526 &
10.748 & 6696 & 9.6 $\times 10^{22}$ &  4.9$\times 10^{13}$ &
-7.6 $\times 10^{2}$\\
CFL4 & 2.526 &
10.709 & 6697& 1.7 $\times 10^{12}$ & $-$ &
-7.8$\times 10^{2}$\\
CFL & 2.526 &
10.729 & 6696 & 1.7 $\times 10^{12}$ & $-$ &
-7.7 $\times 10^{2}$  \\
\end{tabular}
\end{ruledtabular}
\end{center}
\end{table}
The bulk viscosity of the CFL-K0 phase is smaller than that of the 
CFL or ungapped phase for temperatures $T\lesssim 10^{11}$K.
Consequently, bulk viscosity damping timescales are much
longer than the typical timescale for gravitational wave growth,
rendering  
 bulk viscosity  irrelevant to 
\r-mode damping in cooling neutron stars. The damping timescale 
from shear viscosity in the CFL-K0 phase becomes small enough to 
effectively damp the \r-mode as temperatures fall to $T\sim 5\times 10^{10}$K. 
However, this is still dominated by the phonons, since the kaon shear viscosity
is much smaller. The phonon mean free path increases with decreasing $T$, 
therefore, phonon shearing collisions occur less often. 
The mean free path $\lambda_\phi$ is estimated from the kinetic theory
relation  
$\lambda_\phi\sim \eta/(n_\phi p_\phi)$ using phonon density $n_\phi$ and
thermal momentum $p_\phi$, see
Refs.~\cite{Rupak:2007vp,Jaikumar:2008kh,Alford:2009jm}.  
In Fig.~\ref{fig:Viscosities}, we show the mean free path of the
phonons and kaons.
A larger value of the
parameter $C$ corresponds to a larger cross-section for shearing collisions
and therefore yields a smaller mean free path at the same temperature. It can
be seen from Fig.~\ref{fig:Viscosities} that kaon shear viscosity is important
in the temperature range $10^8$K $\lesssim T \lesssim  10^{10}$K, where phonon shear
viscosity is no longer important.

Finally, we compute the critical frequency curve for the CFL-K0 phase 
according to Eq.(\ref{criteqn}). We notice two ``dips'' in the critical
frequency curve: the first one at $T\sim 8\times 10^9$K, when the phonon mean
free path $\lambda_\phi$ becomes larger than the stellar radius, and the second
at $T\sim 2.5 \times 10^8$ when the
kaon mean free path becomes larger. This second dip is sensitive to the
value of $C$, appearing earlier for smaller $C$. In order to obtain
a smooth behaviour for the critical frequency curve, we use for
convenience a sharply peaked cutoff function
$f(\lambda_\phi)=[1-\tanh(\lambda_\phi/{\rm km} -5)]/2$ centered
around a $5$ km stellar radial distance. 
The maximum stable frequency accommodated by the CFL-K0 phase falls 
 well below the observed bound on rotation rates in LMXBs.

\section{\label{sec:summary} Conclusions}

\r-mode oscillations of compact stars in the kaon-condensed CFL phase
are considered. The mode frequency is almost exactly the same as that
for a CFL star~\cite{Jaikumar:2008kh} since the equation of state for the CFL-K0 phase 
only differs at ${\cal O}(m_S^4)$.  The mode growth and viscous damping timescales
in the temperature range  $10^8$K $\lesssim T\lesssim
10^{11}$K is studied based on the dominant bulk and shear viscosity contributions.
Compared to the pure CFL phase, the bulk
viscosity in CFL-K0 is smaller and does not play a significant role. 
At temperatures $T\gtrsim 10^{10}$K shear viscosity
associated with phonon scattering is important and determines the
temperature dependence of the critical frequency curve. At 
lower temperatures, the phonon mean free path exceeds the star radius, so that the new mechanism for
shear viscosity associated with kaon condensation~\cite{Alford:2009jm} provides the
dominant dissipation.     
In comparison to pure CFL stars,
CFL-K0 stars are more stable against the \r-mode instability 
 at the lower end of the temperature range considered. 
However, 
given the observed LMXB spin rates, 
the 
critical frequency curve predicts that even pure CFL-K0 stars are unlikely to exist.
Just as in the neutron matter case, viscous damping
just beneath the core-crust interface (Ekman layer) appears to be necessary to provide a
large enough stable rotation rate consistent with LMXB data~\cite{Bildsten:1999zn,*Levin:2000vq}.

The main result of the current work is that given the known dissipative mechanism in
CFL or CFL-K0 phase, a pure quark star with these phases is ruled out by 
observed LMXB spin rates. Mutual friction associated
with kaon-vortex scattering could provide an additional dissipative
mechanism, but the critical frequency
curves calculated here provide a lower bound. Future related work
should investigate hybrid stars with a  
CFL-K0 core. The role of the crust-core interface as well as the
quark-hadronic matter layer should be key ingredients in obtaining 
constraints on the persistence of gravitational waves from stars 
containing superfluid quark matter in their interior.


%

\end{document}